%% file: CuF2_H2O__pyz_2_paper.tex
\documentclass[aps,prl,showpacs,superscriptaddress,10pt,twocolumn]{revtex4-1}

\usepackage{amsmath}

\usepackage{url}
\usepackage[load-configurations=abbreviations,detect-display-math=true]{siunitx}
\usepackage{graphicx}

\def \musr {$\mu^+$SR}

\begin{document}

\title{Evolution of magnetic interactions in a pressure-induced Jahn-Teller driven magnetic dimensionality switch}

\author{S. Ghannadzadeh}
\email{s.ghannadzadeh1@physics.ox.ac.uk}

\author{J. S. M\"{o}ller}
\email{j.moeller1@physics.ox.ac.uk\ for \musr\ details.}
\author{P. A. Goddard}
\affiliation{Clarendon Laboratory, Department of Physics, University of Oxford, Parks Road, Oxford, OX1 3PU, UK}

\author{T. Lancaster} 
\affiliation{Department of Physics, Durham University, South Road, Durham, DH1 3LE, UK}

\author{F. Xiao}
\affiliation{Department of Physics, Durham University, South Road, Durham, DH1 3LE, UK}

\author{S. J. Blundell}
\affiliation{Clarendon Laboratory, Department of Physics, University of Oxford, Parks Road, Oxford, OX1 3PU, UK}

\author{A. Maisuradze}
\author{R. Khasanov}
\affiliation{Laboratory for Muon Spin Spectroscopy, Paul Scherrer Institut, CH-5232 Villigen PSI, Switzerland}

\author{J. L. Manson}
\affiliation{Department of Chemistry and Biochemistry, Eastern Washington University,
Cheney, WA 99004, USA}

\author{S. W. Tozer}
\author{D. Graf}
\affiliation{National High Magnetic Field Laboratory, Florida State University, Tallahassee, FL 32310, USA}

\author{J. A. Schlueter}
\affiliation{Materials Science Division, Argonne National Laboratory, Argonne, IL 60439, USA}

\begin{abstract}
We present the results of high-field magnetization and muon-spin relaxation measurements on the coordination polymer CuF$_2$(H$_2$O)$_2$(pyrazine)  in pressures up to \SI{22.5}{kbar}. We observe a transition from a quasi-two-dimensional to a quasi-one-dimensional antiferromagnetic phase at \SI{9.1}{kbar}, driven by a rotation of the Jahn-Teller axis. Long-range antiferromagnetic ordering is seen in both regimes, as well as a phase separation in the critical pressure region. The magnetic dimensionality switching as pressure is increased is accompanied by a halving of the primary magnetic exchange energy $J$ and a fivefold decrease in the ordering temperature $T_\text{N}$. $J$ decreases gradually with pressure in the two-dimensional phase, and then increases in the one-dimensional regime. We relate both effects to the changes in the crystal structure with applied pressure.
\end{abstract}

\maketitle

\input{intro}
\input{magnetisation}

\input{musr}

\input{discussion}

\begin{acknowledgments}
This work is supported by EPSRC (UK). A portion of this work was performed at the National High Magnetic Field Laboratory, which is supported by NSF Cooperative Agreement No. DMR-0654118, the State of Florida, and the DOE.  The muon experiment was performed on the GPD instrument at the Swiss Muon Source, Paul Scherrer Institut, Switzerland. Work supported by Argonne, a DOE Office of Science laboratory, operated under Contract No. DE-AC02-06CH11357.  DG and SWT are supported by DOE/NNSA Grant No. DE-FG52-10NA29659. The work at EWU was supported by NSF Grant No. DMR-1005825.
\end{acknowledgments}
\bibliographystyle{apsrev4-1}
\bibliography{database}

\end{document}

%% file: intro.tex
Pressure plays a central role in the exploration of physical phenomena, for example in metal-insulator transitions \cite{Okabe2011,Irizawa2011} and superconductivity \cite{Graf2012,Park2006}. This is because it allows a controlled adjustment of structural parameters, such that their influence on the electronic and magnetic properties can be determined. Pressure can also affect transition-metal co-ordination polymers, which are important candidates for the future development of purpose-engineered magnetic materials \cite{Blundell2007}. A portion of such polymeric magnets are based on hetero-ligand Jahn-Teller (JT) \cite{Jahn1937,*Falvello1997} active metal centers, where the transition metal ion sits in a position of octahedral symmetry, surrounded by an asymmetric ligand environment.  In such systems each trans-coordinated ligand provides an additional degree of freedom on the JT-axis, meaning that small perturbations of the metal-ligand environment can be enough to rotate the JT-axis and radically modify the material properties. Pressure presents an ideal method of systematically achieving such perturbations, due to the relative softness and high compressibility of the organic framework. It has recently been shown that it is possible to select the magnetic dimensionality of polymeric magnets at the synthesis stage \cite{Goddard2012a}. In this Letter we will show that pressure can be used to operate a Jahn-Teller driven magnetic dimensionality switch, between a two dimensional and a one-dimensional antiferromagnetic phase, in the extended Cu-based coordination network CuF$_2$(H$_2$O)$_2$(pyz) (pyz = pyrazine, C$_4$H$_4$N$_2$). In this compound, hydrogen-bonding interactions, a JT-active metal center, and three different trans-coordinated ligands promote significant pressure-induced structural transitions. Together, these features enable pressure-induced perturbation of the magnetic exchange, allowing the modification of the system's magnetic dimensionality and providing the ability to tune the primary exchange energy $J$. 

\begin{figure}[!tb]
\includegraphics[width=1 \columnwidth]{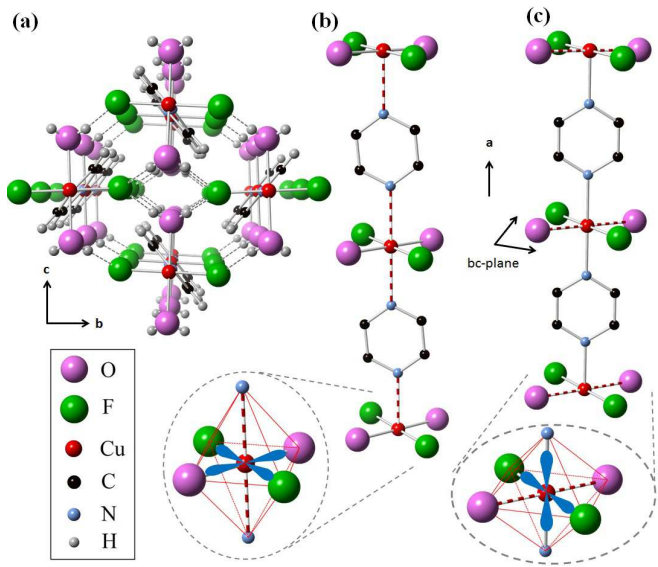}
	\caption{\label{fig:Structure} (color online) Crystal structure of CuF$_2$(H$_2$O)$_2$(pyz) \cite{Manson2008,Halder2011}, showing  (a) the 2D hydrogen bonding network at ambient pressure. Hydrogen positions are approximate. Also shown is an isolated Cu--pyz--Cu chain at (b) ambient pressure and (c) \SI{13.9}{kbar}, on the same scale. The JT-axis is shown by the red striped bonds. The zoomed-in regions show the JT octahedron, together with a representation of the magnetic orbitals, which provide the dominant exchange pathways. } 
\end{figure}

CuF$_2$(H$_2$O)$_2$(pyz) has a strongly anisotropic structure, consisting of Cu--pyz--Cu chains along the crystallographic $a$-axis, with the Cu$^\text{2+}$ metal centers also joined together via a two-dimensional hydrogen-bonding lattice in the $bc$-plane, as shown in Fig. \ref{fig:Structure}. The spin--$1/2$ Cu atom sits at the center of a distorted octahedron, surrounded by trans-pairs of Cu--O, Cu--F and Cu--N ligands. At ambient pressure the elongated JT-axis is oriented along the N--Cu--N bonds, implying that the $d_{x^2-y^2}$ orbitals and the directions of largest electronic overlap are directed perpendicular to this axis, and are within the $bc$-plane. This situation is confirmed by pulsed-field magnetization, electron-spin resonance (ESR),  muon-spin relaxation (\musr), and  neutron scattering measurements \cite{Manson2008,Wang2012}, showing that the compound displays highly two-dimensional antiferromagnetic behavior ($J$ = 11.5 K, secondary exchange energy $J_\perp \sim 10^{-4} J$, long-range ordering temperature $T_N$ = \SI{2.6}{K}), with the primary magnetic exchange mediated via the Cu--OH$\cdots$F--Cu spin exchange paths within the $bc$-planes. The crucial role of the H$\cdots$F bonds in mediating exchange has also been investigated through selective isotopic H-substitution \cite{Goddard2008a}.

Recent synchrotron x-ray powder diffraction measurements by \citet{Halder2011} have revealed that CuF$_2$(H$_2$O)$_2$(pyz) undergoes two structural phase transitions as a function of pressure, accompanied by a rotation of the JT-axis and orbital orientations. With the application of pressure, the elongated JT axis was found to change from the Cu--N to the Cu--O bond direction at $P_\mathrm{c} \approx$ \SI{9}{kbar}. This shift was also detected in single-crystal diffraction experiments, albeit at the higher pressure of \SI{18}{kbar} \cite{Prescimone2012}. A further pressure induced change to the Cu--F direction was seen by \citet{Halder2011} at \SI{31}{kbar}.

In this paper we report a profound change in the magnetic properties occurring at \SI{9.1}{kbar} in both single-crystal and powder samples. Our interpretation of this result is consistent with the rotation of the JT-axis observed by \citet{Halder2011}.  At this pressure the magnetic structure switches from quasi-two-dimensional (Q2D) to quasi-one-dimensional (Q1D) due to the reorientation of the $d_{x^2-y^2}$ orbitals, from lying within the hydrogen-bonded $bc$-plane to the $ac$-plane --- thus significantly decreasing the electronic overlap along the Cu-OH$\cdots$F-Cu bonds, while simultaneously increasing the magnetic orbital overlap along the $a$-axis \cite{Halder2011}. This results in the primary exchange being mediated by pyrazine molecules via the Cu--pyz--Cu chain, therefore allowing Q1D antiferromagnetism to develop. To establish this pressure-induced dimensional switching, as well as to track the evolution of magnetic energies, the behavior of CuF$_2$(H$_2$O)$_2$(pyz) as a function of pressure was explored using high-field magnetization and \musr.

\begin{figure*}[!bt]
\includegraphics[width=1 \textwidth]{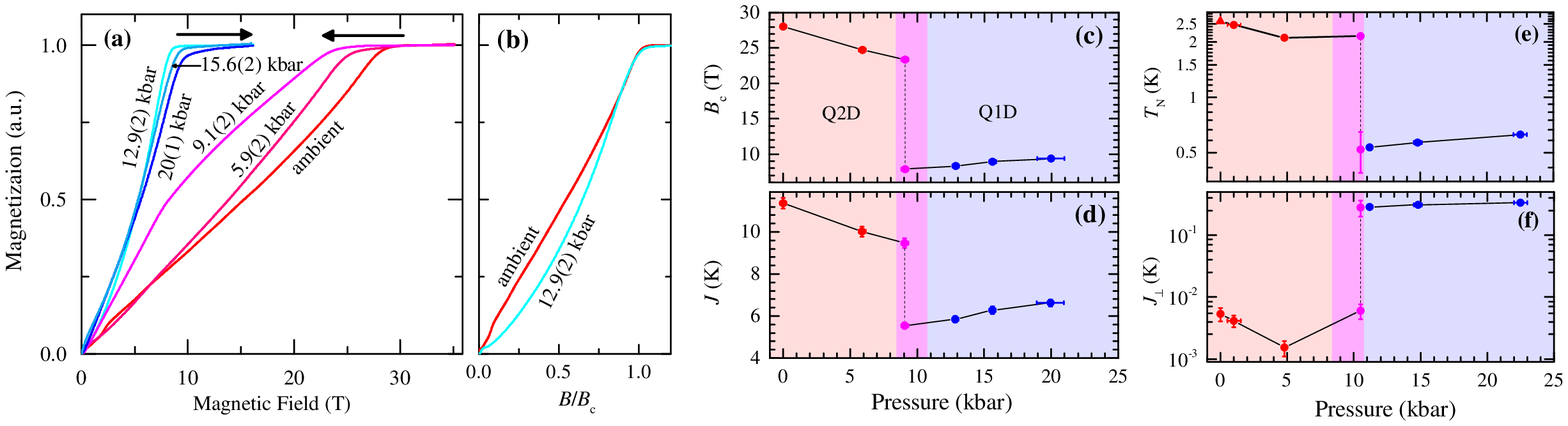}
	\caption{\label{fig:Results}(color online) (a) Magnetization of CuF$_2$(H$_2$O)$_2$(pyz) single crystals with $B\parallel a$-axis, at \SI{1.4}{K} (\SI{2}{K} at \SI{15.6}{kbar}). The arrows represent the change in the saturation field with increasing pressure. (b) $M$ vs $B/B_\mathrm{c}$ at ambient pressure and \SI{12.9}{kbar}, showing the change in curvature.  Also shown is (c) the saturation field $B_\mathrm{c}$,  (d) the deduced primary exchange coupling energy $J$,  (e) the ordering temperature $T_{\rm N}$ deduced from the \musr\ data (see Fig. \ref{fig:muons}), and (f) the secondary exchange $J_\perp$. To extract $J$, we have used values of $g$ obtained from ESR measurements \cite{Prescimone2012}. $T_\mathrm{N}$ at  ambient pressure  (triangle symbol) is from Ref. \cite{Manson2008}. } 
\end{figure*}

%% file: magnetisation.tex
Single crystal magnetization measurements in fields up to \SI{35}{T} were performed at NHMFL, USA, using a novel radio-frequency technique that has recently been shown to be an effective method of obtaining magnetic susceptibility, and hence, magnetization \cite{Ghannadzadeh2011, suppl}. Measurements were performed at \SI{1.4}{K}, with the applied magnetic field parallel to the Cu--pyz--Cu chain direction. A piston cylinder cell was used to achieve the pressure. The pressure was measured in-situ by using the fluorescence of ruby, which showed good hydrostatic conditions \cite{suppl}. Two measurements were performed in the low pressure regime ($P < P_\mathrm{c}$), another at the critical pressure $P_\mathrm{c}$, and a further three in the high pressure region ($P > P_\mathrm{c}$). 

The magnetization, measured across a range of pressures up to \SI{20}{kbar}, is given in Fig. \ref{fig:Results}(a). At ambient pressure the magnetization rises monotonically with a curvature typical of Q2D antiferromagnetic systems \cite{Goddard2012a}, saturating at \SI{28.1}{T} in agreement with previous measurements \cite{Manson2008}. As pressure is applied, we observe a steady reduction in the critical field $B_\mathrm{c}$, down to \SI{23.4}{T} at \SI{9.1}{kbar} [see Fig. \ref{fig:Results}(c)]. Further application of pressure leads to a dramatic drop in $B_\mathrm{c}$ to \SI{8.3}{T}, after which any additional pressure serves to gradually increase the saturation field, in contrast to the behavior seen at lower pressures.

The magnetism in CuF$_2$(H$_2$O)$_2$(pyz) can be described by the Heisenberg model
\begin{equation*}
\mathcal{H}= J \sum_{\langle i,j\rangle_\parallel}\mathbf{S}_i.\mathbf{S}_j + J_\perp \sum_{\langle i,j\rangle_{\perp}}\mathbf{S}_i.\mathbf{S}_j - g\mu_{\mathrm B} B \sum_{i} S_i^z,
\end{equation*}
where in a Q2D (Q1D) phase, $J$ and $J_\perp$ are the magnetic exchange strengths within and normal to the planes (chains), respectively. The first two summations are over the unique ligands through which the exchange coupling takes place, and the last term is the Zeeman splitting term. For a Heisenberg spin--1/2 antiferromagnet (AFM) the form of the magnetization up to saturation strongly reflects the magnetic dimensionality, and Quantum Monte-Carlo simulations have shown that the magnetization becomes strongly concave as the dimensionality is reduced \cite{Goddard2008}. Indeed, looking at Fig. \ref{fig:Results}(b), the magnetization becomes noticeably concave above \SI{9.1}{kbar}, an indication of the reduction in the magnetic dimensionality. Moreover, looking at the \SI{9.1}{kbar} data, we discover a noticeable change in the magnetization gradient at a field (\SI{7.9}{T}) close to the critical field of the higher pressure phase. The curvature of the magnetization below this field is similar to that of the higher pressure regime, while the curvature above it is closer to that seen at low pressures [see Fig. \ref{fig:Results}(a)]. This implies a degree of phase separation, with coexisting volumes of the sample in the high and low pressure phases.

The fact that both the phase separation and the change in curvature occur at the same pressure as the JT-axis shift \cite{Halder2011} leads us to conclude that the dramatic drop in $B_\mathrm{c}$ above \SI{9.1}{kbar} is due to the proposed transition from the Q2D to a Q1D antiferromagnetic phase. In a strongly anisotropic system ($J_\perp \ll J$), we can express the saturation field as $B_\mathrm{c} \approx n J / g \mu_{\mathrm B}$, where $n$  is the number of nearest neighbor exchange ligands of interaction strength $J$ ($n=4$ for Q2D, 2 for Q1D), and $g$ is the $g$-factor \cite{Goddard2008}. The sudden change in $B_\mathrm{c}$ can thus be understood as a combination of the change in $n$ from 4 to 2 (i.e. a Q2D to Q1D transition), and a simultaneous change in $J$.

The extracted values of $J$ are given in Fig. \ref{fig:Results}(d). We find $J\approx \SI{11.4}{K}$ at ambient pressure, confirming previous pulsed-field measurements \cite{Manson2008}. Under application of moderate pressures, $J$ begins to decrease gradually at a rate of $0.21 \pm \SI{0.01}{K\per kbar}$, eventually reaching \SI{9.5}{K} at 9.1~kbar. A further increase in pressure leads to a substantial reduction in $J$, down to \SI{5.5}{K}, caused by the switching of the main exchange pathways to bonds with weaker effective nearest-neighbor interactions as the system undergoes a change in magnetic dimensionality. Within the high pressure region, any further application of pressure results in a steady increase in $J$ at the rate of $ 0.1 \pm \SI{0.01}{K\per kbar}$.

%% file: musr.tex
Further support for the dimensionality switch comes from \musr\ measurements which probe the magnetic ordering of a system and allow one to extract $T_{\rm N}$ \cite{Blundell1999}. Powder sample \musr\ measurements were carried out at the Swiss Muon Source, with detailed temperature scans taken at six different pressures using a piston cylinder cell, which showed good hydrostaticity \cite{suppl}.  Example \musr\ spectra are given in Figs. \ref{fig:muons}(a-c). The \musr\ data were fitted to
\begin{align*}
A(t) = \sum_i A_i e^{-\lambda_i t} &\cos(2\pi \nu_i t + \phi_i) + \sum_i  A_i e^{-\lambda_i t} + A_\text{b}(t),
\end{align*}
where $\nu_i$ are the dominant oscillation frequencies due to long-range magnetic order of the sample, $\phi_i$ is a constant phase,  the exponential terms account for residual magnetic dynamics in the sample, and $A_\text{b}(t)$ accounts for the background from the pressure cell \cite{suppl}.

The evolution of the fitted oscillation frequencies is given in Figs.~\ref{fig:muons}(d-f). Within the low-pressure regime, we observe oscillations at three frequencies below a critical temperature, which at approximately 1~kbar is \SI{2.48(1)}{K}. This is in agreement with previous ambient pressure measurements which found oscillations at three frequencies below \SI{2.59(1)}{K} \cite{Manson2008}. In this regime, $T_\mathrm{N}$ is found to decrease  with application of pressure [see Fig.~\ref{fig:Results}(e)].

At 11.2~kbar and above there is no discontinuous change in the \musr\ spectra above 1~K. However, oscillations at a single frequency appear below 630~mK [Figs. \ref{fig:muons}(c,f)], providing evidence for long-range magnetic order in the high pressure regime. $T_{\mathrm{N}}$ is found to increase with increasing pressure, in accordance with the behavior of the dominant exchange $J$, and the change in $T_{\mathrm{N}}$  is found to be reversible upon reduction of pressure from 22.5~kbar to 11.2~kbar. Compared to a Q2D system, a Q1D system should be expected to have a much reduced long-range ordering temperature owing to the increased influence of fluctuations in one dimension. Therefore,  the large decrease in $T_{\mathrm{N}}$ that we observe in the critical pressure region [Fig.~\ref{fig:Results}(e)]  is consistent with a transition to Q1D magnetic behavior at high pressures. Also note the different temperature dependence of the oscillation frequencies in the Q1D and Q2D phases.

%% file: discussion.tex
The combination of a substantial change in the \musr\ frequencies and ordering temperature $T_{\rm N}$, as well as a dramatic reduction in the critical field $B_\mathrm{c}$, provides unambiguous evidence and direct confirmation of the proposed phase transition from Q2D to Q1D antiferromagnetism. We emphasize that the critical pressure region of \SI{9.1}{kbar} is in agreement with the critical pressure of the JT-axis change seen in powder-sample structural measurements \cite{Halder2011}.  This dimensionality crossover is further supported by an ESR study, which observed reorientation of the magnetic d-orbitals at the pressure-induced JT-axis shift \cite{Prescimone2012}.

Given the values of $J$ and $T_{\rm N}$, we can estimate the secondary exchange energy $J_\perp$  by using the empirical relations $|J_\perp| = |J| \exp{ \left( 2.43 - 0.88 \lambda \right)}$ for the Q2D phase, and $|J_\perp| = 1.073\,T_\mathrm{N} \left( \ln{\lambda} + 0.5 \ln{\ln{\lambda}} \right)^{-1/2}$ for the Q1D phase, where $\lambda = 2.6\,J/T_\mathrm{N}$ \cite{Yasuda2005}. The calculated values of $J_\perp$ are shown in Fig. \ref{fig:Results}(f). Within the low pressure Q2D phase we find $J_\perp$ to be of the order of \SI{1}{mK}. In the Q1D regime, $J_\perp$ is mediated by the portion of the $d_{x^2-y^2}$ orbitals that still lie within the $bc$-plane, and this leads to a jump in $J_\perp$ by two orders of magnitude to approximately $\SI{0.3}{K}$ as we move over to the Q1D phase. 

\begin{figure}[!tb]
\center
\includegraphics[width=1\columnwidth]{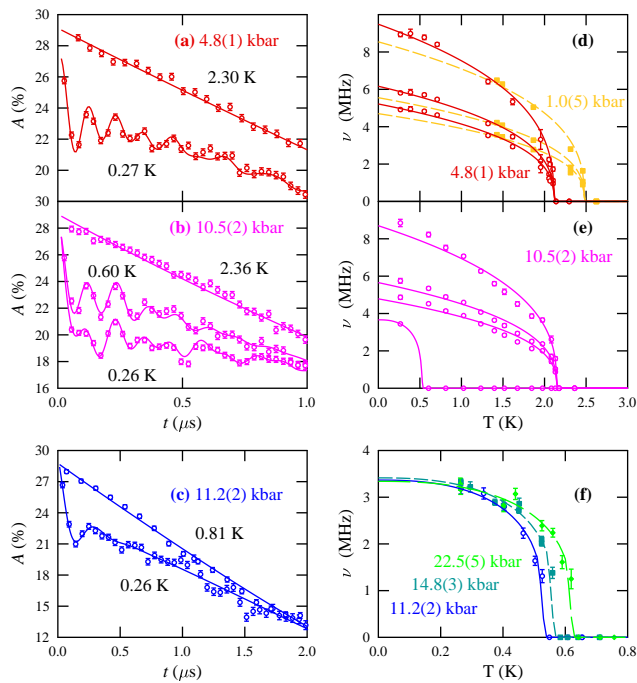}
\caption{\label{fig:muons}(color online) Time evolution of the $\mu^+$ decay asymmetry $A(t)$ in (a-b) the low pressure regime,  and (c) the high pressure phase. Also shown are (d-f) the fitted oscillation frequencies $\nu(T)$. Three frequencies were dominant in the low pressure regime, while the higher pressure phase was well described by a single frequency. The lines in (d-f) represent fits to $\nu(T)=\nu_0 (1-(T/T_{\rm N})^\alpha)^\beta$, where $\alpha$ and $\beta$ were fixed to  1 (2.5) and 0.36 (0.3) in the Q2D (Q1D) phases, respectively.}
\end{figure}

Quantum fluctuations significantly reduce the size of the ordered moment in low-dimensional AFMs. Using a mean-field approximation, we can estimate the zero-field moment of a Q1D AFM from $m_0=2.034 \sqrt{J_{\perp}/J}\, \mu_{\rm B}$ \cite{Schulz1996}. The value of $J_{\perp}/J$ remains approximately constant within the Q1D phase, yielding a value of $m_0=0.46 \pm 0.01\, \mu_{\rm B}$. This compares with $0.5\, \mu_{\rm B}$ and $0.072\, \mu_{\rm B}$ in the Q1D AFMs KCuF$_3$ and Sr$_2$CuO$_3$ \cite{Schulz1996}. We expect nearly the full Cu moment of $1\, \mu_{\rm B}$ in the Q2D phase \cite{suppl}.

The trends in $J$ [Fig. \ref{fig:Results}(d)] can be explained in terms of the effect of pressure on the crystal structure and the magnetic exchange ligands. The decrease seen in $J$ within the Q2D phase is due to the increase in the $\beta$-angle between the $a$ and $c$ axes \cite{Halder2011}, which leads to a misalignment of Cu--pyz--Cu chains and causes them to begin to slide past each other \cite{Musfeldt2011}. This serves to severely disrupt the H$\cdots$F bonded network, decreasing the efficiency of magnetic coupling along the Cu--OH$\cdots$F--Cu exchange pathways  and thus reducing $J$.

The steady increase in $J$ with applied pressure within the Q1D regime  is because unlike the soft H$\cdots$F bonded layers, the Cu--N bonds and the pyrazine rings are relatively resilient, and so no major distortions are seen within the Cu--pyz--Cu chain in the Q1D regime  \cite{Halder2011,Musfeldt2011}. This resiliency, together with the decrease in the chain length with applied pressure \cite{Halder2011}, enhances the magnetic d-orbital density overlap along the chain and results in the increase in $J$.  

The \musr\ data provide evidence for phase separation at 10.5~kbar, where a fourth frequency appears at low temperatures and the other three frequencies remain present up to 2.15~K [see Fig. ~\ref{fig:muons}(e)]. The magnitude of the fourth frequency is consistent with those found for the Q1D ordering at 11.2~kbar and above, and the other three frequencies are consistent with those found for the Q2D ordering below 10~kbar.  While we can not completely rule out the effects of non-hydrostaticity \footnote{The hydrostaticity is estimated to be $\pm \SI{0.2}{kbar}$ for the \SI{10.5}{kbar} \musr\ experiment. We also estimate an upper limit of \SI{+- 0.7}{kbar} for the \SI{9.1}{kbar} magnetization measurement \cite{suppl}.}, these observations in both powder \musr\ and single crystal magnetization measurements, as well as in the separate ESR study \cite{Prescimone2012}, seem to point towards the presence of an intrinsic phase separation in the critical pressure region, reminiscent of the phase separation frequently observed close to some pressure-induced quantum phase transitions \cite{Pfleiderer2005}.

We observe that in CuF$_2$(H$_2$O)$_2$(pyz), the presence of three different ligands in the Cu-centered octahedra effectively introduced several degrees of freedom in the JT-axis orientation. This has allowed a sequential change in the direction of the JT-axis through structural perturbations, due to the different ligand strengths. It is interesting to compare this material to Cu(NO$_3$)$_2$(pyz), which is similarly based on a JT-active Cu atom sitting in a distorted octahedron, but only surrounded by two unique ligands \cite{suppl}. This system is known to be a Q1D AFM at ambient pressure, with the magnetic orbitals lying within the Cu-pyz-Cu chain direction and the JT-axis along the Cu--O bonds \cite{Lancaster2006}. Unlike CuF$_2$(H$_2$O)$_2$(pyz), no JT-switching or orbital reorientation is seen upon application of pressure (up to \SI{11.6}{kbar}) in this compound, due to the angle of the O--Cu--O bonds in the octahedra being significantly different from \SI{90}{\degree} \cite{suppl}. Such a mismatch with the symmetry of the $d_{x^2-y^2}$ orbitals would considerably increase the energy cost of an orbital reorientation. Furthermore, in CuF$_2$(H$_2$O)$_2$(pyz), the change in JT-axis from Cu--N to Cu--O, to Cu--F, leads to a sequential decrease in the unit cell volume with increasing pressure. In contrast, a rotation of JT distortion axis from Cu--O to Cu--N in Cu(NO$_3$)$_2$(pyz) could potentially lead to an increase in the unit cell volume, which is energetically unfavorable.

In summary, we have shown that pressure can be used to induce a Jahn-Teller driven magnetic dimensionality switch in CuF$_2$(H$_2$O)$_2$(pyz), due to a change in the primary magnetic exchange pathways.  Thus pressure can be used to take control of the magnetic properties of metal-polymeric systems, giving us the ability to tune the interaction strengths and select the magnetic dimensionality via external perturbation. This work emphasizes the powerful role that pressure can play in exploring new materials and phases, especially in hetero-ligand complexes with active Jahn-Teller centers.

%% file: CuF2_H2O__pyz_2_paper.bbl
\begin{thebibliography}{23}%
\makeatletter
\providecommand \@ifxundefined [1]{%
 \@ifx{#1\undefined}
}%
\providecommand \@ifnum [1]{%
 \ifnum #1\expandafter \@firstoftwo
 \else \expandafter \@secondoftwo
 \fi
}%
\providecommand \@ifx [1]{%
 \ifx #1\expandafter \@firstoftwo
 \else \expandafter \@secondoftwo
 \fi
}%
\providecommand \natexlab [1]{#1}%
\providecommand \enquote  [1]{``#1''}%
\providecommand \bibnamefont  [1]{#1}%
\providecommand \bibfnamefont [1]{#1}%
\providecommand \citenamefont [1]{#1}%
\providecommand \href@noop [0]{\@secondoftwo}%
\providecommand \href [0]{\begingroup \@sanitize@url \@href}%
\providecommand \@href[1]{\@@startlink{#1}\@@href}%
\providecommand \@@href[1]{\endgroup#1\@@endlink}%
\providecommand \@sanitize@url [0]{\catcode `\\12\catcode `\$12\catcode
  `\&12\catcode `\#12\catcode `\^12\catcode `\_12\catcode `\%12\relax}%
\providecommand \@@startlink[1]{}%
\providecommand \@@endlink[0]{}%
\providecommand \url  [0]{\begingroup\@sanitize@url \@url }%
\providecommand \@url [1]{\endgroup\@href {#1}{\urlprefix }}%
\providecommand \urlprefix  [0]{URL }%
\providecommand \Eprint [0]{\href }%
\providecommand \doibase [0]{http://dx.doi.org/}%
\providecommand \selectlanguage [0]{\@gobble}%
\providecommand \bibinfo  [0]{\@secondoftwo}%
\providecommand \bibfield  [0]{\@secondoftwo}%
\providecommand \translation [1]{[#1]}%
\providecommand \BibitemOpen [0]{}%
\providecommand \bibitemStop [0]{}%
\providecommand \bibitemNoStop [0]{.\EOS\space}%
\providecommand \EOS [0]{\spacefactor3000\relax}%
\providecommand \BibitemShut  [1]{\csname bibitem#1\endcsname}%
\let\auto@bib@innerbib\@empty
\bibitem [{\citenamefont {Okabe}\ \emph {et~al.}(2011)\citenamefont {Okabe}
  \emph {et~al.}}]{Okabe2011}%
  \BibitemOpen
  \bibfield  {author} {\bibinfo {author} {\bibfnamefont {H.}~\bibnamefont
  {Okabe}} \emph {et~al.},\ }\href {\doibase 10.1103/PhysRevB.84.115127}
  {\bibfield  {journal} {\bibinfo  {journal} {Phys. Rev. B}\ }\textbf {\bibinfo
  {volume} {84}},\ \bibinfo {pages} {115127} (\bibinfo {year}
  {2011})}\BibitemShut {NoStop}%
\bibitem [{\citenamefont {Irizawa}\ \emph {et~al.}(2011)\citenamefont {Irizawa}
  \emph {et~al.}}]{Irizawa2011}%
  \BibitemOpen
  \bibfield  {author} {\bibinfo {author} {\bibfnamefont {A.}~\bibnamefont
  {Irizawa}} \emph {et~al.},\ }\href {\doibase 10.1103/PhysRevB.84.235116}
  {\bibfield  {journal} {\bibinfo  {journal} {Phys. Rev. B}\ }\textbf {\bibinfo
  {volume} {84}},\ \bibinfo {pages} {235116} (\bibinfo {year}
  {2011})}\BibitemShut {NoStop}%
\bibitem [{\citenamefont {Graf}\ \emph {et~al.}(2012)\citenamefont {Graf} \emph
  {et~al.}}]{Graf2012}%
  \BibitemOpen
  \bibfield  {author} {\bibinfo {author} {\bibfnamefont {D.}~\bibnamefont
  {Graf}} \emph {et~al.},\ }\href {\doibase 10.1103/PhysRevB.85.134503}
  {\bibfield  {journal} {\bibinfo  {journal} {Phys. Rev. B}\ }\textbf {\bibinfo
  {volume} {85}},\ \bibinfo {pages} {134503} (\bibinfo {year}
  {2012})}\BibitemShut {NoStop}%
\bibitem [{\citenamefont {Park}\ \emph {et~al.}(2006)\citenamefont {Park} \emph
  {et~al.}}]{Park2006}%
  \BibitemOpen
  \bibfield  {author} {\bibinfo {author} {\bibfnamefont {T.}~\bibnamefont
  {Park}} \emph {et~al.},\ }\href {http://dx.doi.org/10.1038/nature04571}
  {\bibfield  {journal} {\bibinfo  {journal} {Nature}\ }\textbf {\bibinfo
  {volume} {440}},\ \bibinfo {pages} {65} (\bibinfo {year} {2006})}\BibitemShut
  {NoStop}%
\bibitem [{\citenamefont {Blundell}(2007)}]{Blundell2007}%
  \BibitemOpen
  \bibfield  {author} {\bibinfo {author} {\bibfnamefont {S.~J.}\ \bibnamefont
  {Blundell}},\ }\href {\doibase 10.1080/00107510801967415} {\bibfield
  {journal} {\bibinfo  {journal} {Contemporary Physics}\ }\textbf {\bibinfo
  {volume} {48}},\ \bibinfo {pages} {275} (\bibinfo {year} {2007})}\BibitemShut
  {NoStop}%
\bibitem [{\citenamefont {Jahn}\ and\ \citenamefont {Teller}(1937)}]{Jahn1937}%
  \BibitemOpen
  \bibfield  {author} {\bibinfo {author} {\bibfnamefont {H.~A.}\ \bibnamefont
  {Jahn}}\ and\ \bibinfo {author} {\bibfnamefont {E.}~\bibnamefont {Teller}},\
  }\href {\doibase 10.1098/rspa.1937.0142} {\bibfield  {journal} {\bibinfo
  {journal} {Proc. R. Soc. Lond.}\ }\textbf {\bibinfo {volume} {161}},\
  \bibinfo {pages} {220} (\bibinfo {year} {1937})}\BibitemShut {NoStop}%
\bibitem [{\citenamefont {Falvello}(1997)}]{Falvello1997}%
  \BibitemOpen
  \bibfield  {author} {\bibinfo {author} {\bibfnamefont {L.~R.}\ \bibnamefont
  {Falvello}},\ }\href {\doibase 10.1039/A703548I} {\bibfield  {journal}
  {\bibinfo  {journal} {J. Chem. Soc.{,} Dalton Trans.}\ }\textbf {\bibinfo
  {volume} {23}},\ \bibinfo {pages} {4463} (\bibinfo {year}
  {1997})}\BibitemShut {NoStop}%
\bibitem [{\citenamefont {Goddard}\ \emph {et~al.}(2012)\citenamefont {Goddard}
  \emph {et~al.}}]{Goddard2012a}%
  \BibitemOpen
  \bibfield  {author} {\bibinfo {author} {\bibfnamefont {P.~A.}\ \bibnamefont
  {Goddard}} \emph {et~al.},\ }\href {\doibase 10.1103/PhysRevLett.108.077208}
  {\bibfield  {journal} {\bibinfo  {journal} {Phys. Rev. Lett.}\ }\textbf
  {\bibinfo {volume} {108}},\ \bibinfo {pages} {077208} (\bibinfo {year}
  {2012})}\BibitemShut {NoStop}%
\bibitem [{\citenamefont {Manson}\ \emph {et~al.}(2008)\citenamefont {Manson}
  \emph {et~al.}}]{Manson2008}%
  \BibitemOpen
  \bibfield  {author} {\bibinfo {author} {\bibfnamefont {J.~L.}\ \bibnamefont
  {Manson}} \emph {et~al.},\ }\href {\doibase 10.1021/cm8016566} {\bibfield
  {journal} {\bibinfo  {journal} {Chem. Mater.}\ }\textbf {\bibinfo {volume}
  {20}},\ \bibinfo {pages} {7408} (\bibinfo {year} {2008})}\BibitemShut
  {NoStop}%
\bibitem [{\citenamefont {Wang}\ \emph {et~al.}(2012)\citenamefont {Wang} \emph
  {et~al.}}]{Wang2012}%
  \BibitemOpen
  \bibfield  {author} {\bibinfo {author} {\bibfnamefont {C.~H.}\ \bibnamefont
  {Wang}} \emph {et~al.},\ }\href {\doibase 10.1103/PhysRevB.86.064439}
  {\bibfield  {journal} {\bibinfo  {journal} {Phys. Rev. B}\ }\textbf {\bibinfo
  {volume} {86}},\ \bibinfo {pages} {064439} (\bibinfo {year}
  {2012})}\BibitemShut {NoStop}%
\bibitem [{\citenamefont {Goddard}\ \emph
  {et~al.}(2008{\natexlab{a}})\citenamefont {Goddard} \emph
  {et~al.}}]{Goddard2008a}%
  \BibitemOpen
  \bibfield  {author} {\bibinfo {author} {\bibfnamefont {P.~A.}\ \bibnamefont
  {Goddard}} \emph {et~al.},\ }\href {\doibase 10.1103/PhysRevB.78.052408}
  {\bibfield  {journal} {\bibinfo  {journal} {Phys. Rev. B}\ }\textbf {\bibinfo
  {volume} {78}},\ \bibinfo {pages} {052408} (\bibinfo {year}
  {2008}{\natexlab{a}})}\BibitemShut {NoStop}%
\bibitem [{\citenamefont {Halder}\ \emph {et~al.}(2011)\citenamefont {Halder}
  \emph {et~al.}}]{Halder2011}%
  \BibitemOpen
  \bibfield  {author} {\bibinfo {author} {\bibfnamefont {G.~J.}\ \bibnamefont
  {Halder}} \emph {et~al.},\ }\href {\doibase 10.1002/anie.201003380}
  {\bibfield  {journal} {\bibinfo  {journal} {Angew. Chem. Int. Ed.}\ }\textbf
  {\bibinfo {volume} {50}},\ \bibinfo {pages} {419} (\bibinfo {year}
  {2011})}\BibitemShut {NoStop}%
\bibitem [{\citenamefont {Prescimone}\ \emph {et~al.}(2012)\citenamefont
  {Prescimone} \emph {et~al.}}]{Prescimone2012}%
  \BibitemOpen
  \bibfield  {author} {\bibinfo {author} {\bibfnamefont {A.}~\bibnamefont
  {Prescimone}} \emph {et~al.},\ }\href {\doibase 10.1002/anie.201202367}
  {\bibfield  {journal} {\bibinfo  {journal} {Angew. Chem. Int. Ed.}\ }\textbf
  {\bibinfo {volume} {51}},\ \bibinfo {pages} {7490} (\bibinfo {year}
  {2012})}\BibitemShut {NoStop}%
\bibitem [{\citenamefont {Ghannadzadeh}\ \emph {et~al.}(2011)\citenamefont
  {Ghannadzadeh} \emph {et~al.}}]{Ghannadzadeh2011}%
  \BibitemOpen
  \bibfield  {author} {\bibinfo {author} {\bibfnamefont {S.}~\bibnamefont
  {Ghannadzadeh}} \emph {et~al.},\ }\href {\doibase 10.1063/1.3653395}
  {\bibfield  {journal} {\bibinfo  {journal} {Rev. Sci. Instrum.}\ }\textbf
  {\bibinfo {volume} {82}},\ \bibinfo {eid} {113902} (\bibinfo {year}
  {2011})}\BibitemShut {NoStop}%
\bibitem [{sup()}]{suppl}%
  \BibitemOpen
  \href@noop {} {}\bibinfo {note} {See Supplemental Material at
  [URL].}\BibitemShut {Stop}%
\bibitem [{\citenamefont {Goddard}\ \emph
  {et~al.}(2008{\natexlab{b}})\citenamefont {Goddard} \emph
  {et~al.}}]{Goddard2008}%
  \BibitemOpen
  \bibfield  {author} {\bibinfo {author} {\bibfnamefont {P.~A.}\ \bibnamefont
  {Goddard}} \emph {et~al.},\ }\href
  {http://stacks.iop.org/1367-2630/10/i=8/a=083025} {\bibfield  {journal}
  {\bibinfo  {journal} {New. J. Phys.}\ }\textbf {\bibinfo {volume} {10}},\
  \bibinfo {pages} {083025} (\bibinfo {year} {2008}{\natexlab{b}})}\BibitemShut
  {NoStop}%
\bibitem [{\citenamefont {Blundell}(1999)}]{Blundell1999}%
  \BibitemOpen
  \bibfield  {author} {\bibinfo {author} {\bibfnamefont {S.~J.}\ \bibnamefont
  {Blundell}},\ }\href {\doibase 10.1080/001075199181521} {\bibfield  {journal}
  {\bibinfo  {journal} {Contemporary Physics}\ }\textbf {\bibinfo {volume}
  {40}},\ \bibinfo {pages} {175} (\bibinfo {year} {1999})}\BibitemShut
  {NoStop}%
\bibitem [{\citenamefont {Yasuda}\ \emph {et~al.}(2005)\citenamefont {Yasuda}
  \emph {et~al.}}]{Yasuda2005}%
  \BibitemOpen
  \bibfield  {author} {\bibinfo {author} {\bibfnamefont {C.}~\bibnamefont
  {Yasuda}} \emph {et~al.},\ }\href {\doibase 10.1103/PhysRevLett.94.217201}
  {\bibfield  {journal} {\bibinfo  {journal} {Phys. Rev. Lett.}\ }\textbf
  {\bibinfo {volume} {94}},\ \bibinfo {pages} {217201} (\bibinfo {year}
  {2005})}\BibitemShut {NoStop}%
\bibitem [{\citenamefont {Schulz}(1996)}]{Schulz1996}%
  \BibitemOpen
  \bibfield  {author} {\bibinfo {author} {\bibfnamefont {H.~J.}\ \bibnamefont
  {Schulz}},\ }\href {\doibase 10.1103/PhysRevLett.77.2790} {\bibfield
  {journal} {\bibinfo  {journal} {Phys. Rev. Lett.}\ }\textbf {\bibinfo
  {volume} {77}},\ \bibinfo {pages} {2790} (\bibinfo {year}
  {1996})}\BibitemShut {NoStop}%
\bibitem [{\citenamefont {Musfeldt}\ \emph {et~al.}(2011)\citenamefont
  {Musfeldt} \emph {et~al.}}]{Musfeldt2011}%
  \BibitemOpen
  \bibfield  {author} {\bibinfo {author} {\bibfnamefont {J.~L.}\ \bibnamefont
  {Musfeldt}} \emph {et~al.},\ }\href {\doibase 10.1021/ic2008039} {\bibfield
  {journal} {\bibinfo  {journal} {Inorg. Chem.}\ }\textbf {\bibinfo {volume}
  {50}},\ \bibinfo {pages} {6347} (\bibinfo {year} {2011})}\BibitemShut
  {NoStop}%
\bibitem [{Note1()}]{Note1}%
  \BibitemOpen
  \bibinfo {note} {The hydrostaticity is estimated to be $\pm \SI {0.2}{kbar}$
  for the \SI {10.5}{kbar} $\mu ^+$SR\ experiment. We also estimate an upper
  limit of \SI {+- 0.7}{kbar} for the \SI {9.1}{kbar} magnetization measurement
  \cite {suppl}.}\BibitemShut {Stop}%
\bibitem [{\citenamefont {Pfleiderer}(2005)}]{Pfleiderer2005}%
  \BibitemOpen
  \bibfield  {author} {\bibinfo {author} {\bibfnamefont {C.}~\bibnamefont
  {Pfleiderer}},\ }\href {http://stacks.iop.org/0953-8984/17/i=11/a=031}
  {\bibfield  {journal} {\bibinfo  {journal} {J. Phys.: Condens. Matter}\
  }\textbf {\bibinfo {volume} {17}},\ \bibinfo {pages} {S987} (\bibinfo {year}
  {2005})}\BibitemShut {NoStop}%
\bibitem [{\citenamefont {Lancaster}\ \emph {et~al.}(2006)\citenamefont
  {Lancaster} \emph {et~al.}}]{Lancaster2006}%
  \BibitemOpen
  \bibfield  {author} {\bibinfo {author} {\bibfnamefont {T.}~\bibnamefont
  {Lancaster}} \emph {et~al.},\ }\href {\doibase 10.1103/PhysRevB.73.020410}
  {\bibfield  {journal} {\bibinfo  {journal} {Phys. Rev. B}\ }\textbf {\bibinfo
  {volume} {73}},\ \bibinfo {pages} {020410} (\bibinfo {year}
  {2006})}\BibitemShut {NoStop}%
\end{thebibliography}%
